\documentclass[%twocolumn,
aps,nofootinbib,
showpacs,showkeys,preprint
tightenlines,preprintnumbers,] {revtex4}

\usepackage{epsf,epsfig,subfigure,graphicx,amsmath,amssymb %,mathtools
}
\usepackage{color}
\newcommand{\dis}[1]{\begin{equation}\begin{split}#1\end{split}\end{equation}}
\newcommand{\ie}{{\it i.e.~}}
\newcommand{\etal}{{\it et al.\,}}

\newcommand{\gev}{\,\textrm{GeV}}
\newcommand{\meV}{\,\mathrm{MeV}}

\newcommand{\eV}{\,\mathrm{eV}}
\newcommand{\Mp}{M_{\rm P}}

\newcommand{\Mg}{{M_{\rm GUT}}}

\newcommand{\Lqcd}{\Lambda_{\rm QCD}}

\newcommand{\USU}{U(1)$_{\rm global}\times \rm SU(3)_{\rm gauge}^2$}
\newcommand{\USW}{U(1)$_{\rm global}\times \rm SU(2)_{\rm gauge}^2$}
\newcommand{\Uanom}{U(1)$_{\rm anom}$}
\newcommand{\alem}{\alpha_{\rm em}}

\newcommand{\UPQ}{U(1)$_{\rm PQ}$}
\newcommand{\Uq}{U(1)$_{\rm q}$}

\newcommand{\UoR}{U(1)$_{\rm R}$}

\def\sw0{{$\sin^2\theta_W^0$}}

\def\Nf2{{\bf N_{[2]}}}

\newcommand{\Z}{{\bf Z}}

\def\SU2Ch{SU(2)$_L\times$SU(2)$_R$}
\def\E6{{\rm E_6}}

\def\EE8{{\rm E_8\times E_8'}}

\def\anti{anti-SU$(5)$}

\def\one{{\bf 1}}

\def\five{{\bf 5}}
\def\ten{{\bf 10}}
\def\tenb{{\overline{\bf 10}}}

\def\fiveb{{\overline{\bf 5}}}

\begin{document}

\draft

\title{\Large\bf Anomalies and parities for quintessential and ultra-light axions}

\author{Jihn E.  Kim }
\address
{Department of Physics, Kyung Hee University, 26 Gyungheedaero, Dongdaemun-Gu, Seoul 02447, Republic of Korea  
}
 
\begin{abstract} 
We discuss the energy scales of the explicit breaking terms of the global symmetries \USW~  needed for the quinessential axion (QA) and the ultra-light axion (ULA). The appropriate scale of QA is about $1.48\times 10^{8\,}\gev$.

\end{abstract}
\maketitle

%%%%%%%%%%%%%%%%%%%%%%%%%%%%%
%%%%%%%%%%%%%%%%%%%%%%%%%%
\section{Introduction}\label{sec:Introduction}

Light pseudoscalar particles are the lampposts to the road leading to physics scales much above their masses. The well-known example is the pion triplet which guided toward the \SU2Ch~chiral symmetry at the strong interaction scale \cite{Nambu61}. Another is the very light axion which hints the intermediate scale \cite{KSVZ1,KNS18}. The scale where the restored symmetry recovered is,
$E\gtrsim f\sim\Lambda^2/m$, for the pseudoscalar mass $m$, where $\Lambda^4$ is a typical energy density contributed by such a pseudoscalar. Before obtaining mass by the energy density  perturbation  $\Lambda^4$, these pseudoscalar degrees are phase degrees $\theta(x)$ of some unitary operators such that they have kinetic energy terms in quantum field theory below the pseudoscalar defining scale $f$. Above the scale $f$, the phase is not a dynamical field, \ie not depending on $x$, but it still represents a phase direction of the global symmetry. Gravity does not respect global symmetries. Global symmetries are destined to be broken by gravity.

If the pseudoscalar mass $m$ is less than 1 MeV, it can decay to two photons and two neutrinos. For these two rare processes, the $\pi^0$ decays give  $\Gamma(\pi^0\to 2\gamma)\simeq ( \alem^2/64\pi^3) (M_{\pi^0}/f_\pi)^2M_{\pi^0}$ (at leading order \cite{Holstein02}) and $1.2\times 10^{-5}\Gamma(\pi^0\to 2\gamma)$ \cite{Shrock79}, respectively. For a pseudoscalar to be still present in our Universe, we require $\Gamma^{-1} > 4.3\times 10^{17}{\rm s} = 1/(1.53\times 10^{-33} \eV)$.
 Using the $f-m$ relation, we get an idea on $m$ from the condition that it survives until now. If $\Lambda$ is the QCD scale 
 \dis{
 \Gamma^{-1}\simeq  \frac{ 64\pi^3\,\Lambda_{\rm QCD}^4}{\alem^2m^4}   \frac{1}{m}>\frac{1}{1.53\times 10^{-33} \eV}
 }
or $m< 65\, \eV$ for $\Lambda_{\rm QCD}=380\,\meV$.\footnote{For the very light axion, the specific coupling reduces it further to 24\,eV \cite{KimRMP10}.} If we take $\Lambda^4$ as the current energy density of the Universe, $(0.003\,\eV)^4$, then pseudoscalars with mass less than $3.3\times 10^{-4}\, \eV$ survives until now. Therefore, very light pseudoscalars attracted a great deal of attention in cosmology. Pseudoscalars for  $m<3.3\times 10^{-4}\, \eV$ can contribute to the current energy density and structure formation. In this context, quinessential axion (QA) \cite{Carroll98,Hill02,KN03} was suggested for dark energy (DE), and ultra-light axion (ULA) \cite{KimPRD16} was suggested for the galactic scale structures \cite{Witten17}. The precursers of these pseudoscalars are `general quintessence', not pinpointing to the pseudoscalars \cite{quinessence}.  
  
For a pseudoscalar created by spontaneous breaking of a global symmetry, there is no pseudoscalar above the defining scale $f$. Below $f$, the fundamental scale for the pseudoscalar decay-interactions is $f$ and hence  $f$  is called the {\it decay constant}. 
On the other hand, if a pseudoscalar is present in the theory, then $f$ is the defining scale of that theory.  In string theory, there are antisymmetric tensor fields which are pseudoscalars, and hence $f$ in this case is the string scale $M_s$. Below the compactification scale, various scales combine to give $f$ above \cite{KNP05} or below \cite{KimPLB17} the string scale, depending on the compactification schemes. In this paper, we will not discuss the ultraviolet completed theories because they depend on the details at the Planck scale physics \cite{ConlonSven16}.

In the effective field theory approach, there are two classes for the explicit breaking scale $\Lambda$: one class in the potential $V$ of spin-0 fields and Yukawa couplings, and the other class the non-abelian gauge anomalies.  Since gravitational interactions always break any global symmetry \cite{Barr92},  we can consider the explicit breaking term at $\Lambda$ due to gravity, of order $\Lambda^{n+4}/\Mp^n$. For this term to be sufficiently small for QA or ULA, theory must possess some symmetry to forbid terms up to the order $\Lambda^{n+3}/\Mp^{n-1}$. In Ref.  \cite{KimPRD16}, discrete symmetries were used to realize this scheme. In this paper, we focus on  the contribution of non-abelian gauge anomalies such as the Peccei-Quinn(PQ) symmetry \cite{PQ77} \USU\, and another global symmetry \USW. 

In Sec.  \ref{sec:Anomaly}, we present the feasibility for the class of gauge anomaly, \USW, to work for DE, and in Subsec. \ref{subsec:Potential} the forbidden and allowed rare processes are discussed.  In Sec. \ref{sec:GravCont}, gravity effects are briefly discussed.  Section \ref{sec:Conclusion} is a conclusion. 

%%%%%%%%%%%%%%%%%%%%%%%%%%
\section{Anomaly breaking}\label{sec:Anomaly}

Non-abelian gauge forces break global symmetries \cite{Belavin75}. In the minimal supersymmetric standard model (MSSM), the gauge coupling runs such that  the weak SU(2)  anomaly may work for DE. We have the following SU(2) gauge coupling at $M_Z$, using $  \alpha_{\rm em}(M_Z)=1/(127.940\pm 0.014)$,
\dis{\frac{1}{\alpha_{2}(M_Z)}= 29.600\pm 0.010. 
}
In the MSSM, there already exists an excellent estimate,  based on the two-loop MSSM running of $\alpha_2$ \cite{Bourilkov15},
%%%%%%%%%%%%
\begin{table}[h!]
\begin{center}
\begin{tabular}{@{}|c| c c c|c|@{}} \hline
Threshold correction [\%] &$M_{SUSY}[\gev],$&  $Mg[\gev],$ &    $1/\alpha_{GUT}$& $\chi^2$\\[0.2em] \colrule &&&&  \\[-1.4em] 
 ~$+1$~  & $10^{3.96\pm 0.10}$  & $10^{15.85\pm 0.03}$ &   $26.74\pm 0.17$ &8.2\%   \\[0.3em]  \\[-1.4em] 
~$\pm 0$~  & $10^{3.45\pm 0.09}$  & $10^{16.02\pm 0.03}$ &   $25.83\pm 0.16$ &8.2\%   \\[0.3em]  \\[-1.4em] 
 ~$-1$~  & $10^{3.02\pm 0.08}$  & $10^{16.16\pm 0.03}$ &   $25.07\pm 0.15$   &9.5\%     \\[0.3em]  \\[-1.4em] 
 ~$-2$~  & $10^{2.78\pm 0.07}$  & $10^{16.25\pm 0.02}$ &   $24.63\pm 0.13$  &25.1\%      \\[0.3em]  \\[-1.4em] 
 ~$-3$~  & $10^{2.60\pm 0.06}$  & $10^{16.31\pm 0.02}$ &   $24.28\pm 0.10$   &68.1\%     \\[0.3em]  \\[-1.4em] 
 ~$-4$~  & $10^{2.42\pm 0.05}$  & $10^{16.38\pm 0.02}$ &   $23.95\pm 0.09$ &138.3\%   \\[0.3em]  \\[-1.4em] 
 ~$-5$~  & $10^{2.26\pm 0.05}$  & $10^{16.44\pm 0.02}$ &   $23.66\pm 0.09$   &235.7\%     \\[-0.1em]
 \hline
\end{tabular}
\end{center}
\caption{Combined fit results - CMS data \cite{CMS1,CMS2} and GUT unification. The threshold correction $\varepsilon_{GUT}$ at $\Mg$ is defined as $\alpha_3(\Mg)=\alpha_{GUT}(1+\varepsilon_{GUT})$. Here, without the threshold correction  the favored SUSY scale turns out to be 2820 GeV. }\label{tabLCMS15}
\end{table}
%%%%%%%%
on the grand unification(GUT) scale as shown in Table \ref{tabLCMS15}. 
For SU($N$) groups, we use the following $\beta$-function with gauge bosons and fermions \cite{Jones74, Caldwell74},
\dis{
\beta=-(\frac{\alpha_s}{4\pi})^2\left(\frac{11}{3}C_2(G)- \frac23\sum_R T(R)-\frac{\alpha_s}{4\pi}\Big(\frac{10}{3}\sum_R   C_2(G)T(R) +2\sum_R {\rm Cashimir}_2({\rm SU}(N))T(R)-\frac{34}{3}(C_2(G))^2\Big)\right)
}
where $\alpha_s$ is the QCD coupling, and
\dis{
C_2({\rm SU}(N))=N,~{\rm Cashimir}_2({\rm SU}(N))=\frac{N^2-1}{2N},~T(R)=\ell(R), \ell({\bf N})=\frac12.
}
 Table \ref{tabLCMS15} was obtained with the following input parameters,
\dis{
 {\rm At }~ M_Z= 91.19\,\gev:~\left\{~\begin{array}{l}\sin^2\theta_W\Big|_{\overline{\rm MS}} = 0.23126 \pm 0.00005,\\[0.3em]
 \alpha_s=  0.1185 \pm 0.0006,\end{array} .\right.
}

 Without considering the threshold corrections at the GUT scale, we obtain   \cite{Bourilkov15},
\begin{eqnarray}
\textrm{MSSM}:&&\left[ \begin{array}{l}
e^{-2\pi/\alpha_2}\Big|_{\Mg}=1.69\times 10^{-81},\\[2em]
\left\{~\begin{array}{l}M_{\rm SUSY}=2820+670-540~\gev,\\[0.3em]
\Mg=(1.065\pm 0.06) \times 10^{16}\gev.\end{array} \right.
\end{array}\right.\\[1em]
\textrm{SM}:&&\left[ \begin{array}{l}
e^{-2\pi/\alpha_2}\Big|_{\Mg}=1.69\times 10^{-131},\\[1em]
\Mg=(1.096\pm 0.06) \times 10^{15}\gev.\end{array}\right. \label{eq:SMefactor}
\label{eq:MSSM}
\end{eqnarray}
Note that the SU(2) coupling does not run with one-loop corrections in the MSSM. In the SM, the SU(2) coupling runs and we obtain a much smaller factor if not considering the threshold corrections. 
If the SU(2) gauge anomaly is responsible for DE, we require the following order for $\Lambda$ in the MSSM,
\dis{
\textrm{MSSM}:~1.69\times 10^{-81}\Lambda^4=(0.003\,\eV)^4\to \Lambda\sim 1.48\times 10^{8}\,\gev,\\
\textrm{SM}:~1.065\times 10^{-131}\Lambda^4=(0.003\,\eV)^4\to \Lambda\sim  5.25\times 10^{20}\,\gev.\label{eq:SU2scale}
}
Thus, in the SM we cannot use the SU(2) gauge anomaly  for the source of DE. But, in the MSSM  the SU(2) gauge anomaly can work for the explicit breaking term for the QA. If it oscillated before the current epoch, then it cannot work for DE but may work for a ULA \cite{KimPRD16}.
 
%%%%%%%%%
\subsection{QCD axion plus QA/ULA}

For DE, we need another global symmetry beyond the PQ symmetry, and the needed discrete symmetry must be very restrictive such that many lower order terms are not allowed. If we consider another U(1) global symmetry for DE, most probably it has a QCD anomaly also. For it to be free of the QCD anomaly, its quantum numbers of quarks must be chosen judiciously.
 In this section, we comment on the SU(2) anomaly toward QA, and the possibility with superpotential terms from string compactification will be interesting but  postponed to a future communication. In Eq. (\ref{eq:SU2scale}), $\Lambda$ for SU(2) anomaly toward QA was calculated as $1.48\times 10^8\,\gev$. Note that, among the terms breaking global symmetries  the QCD anomaly is the largest one, and QA/ULA must be chosen after removing the QCD anomaly part. Let the singlet field housing the QCD axion and QA are $\sigma$ and $\sigma_{\rm quint}\equiv \sigma_q$\ whose VEVs are
\dis{
\langle \sigma\rangle=\frac{f_a}{\sqrt2}\, e^{ia/f_a},~
\langle \sigma_{\rm quint}\rangle=\frac{f_q}{\sqrt2}\, e^{ia_q/f_q}. 
}
Let us introduce two complex scalar fields possessing the following quantum numbers under U(1)$_{\rm PQ}$ and U(1)$_{\rm q}$,
\dis{
\begin{array}{ccc}
&~~\sigma~~ &\sigma_{\rm quint}\\
{\rm U(1)_{PQ} : }& 1 & \Gamma_2\\
{\rm U(1)_{q} : }& \Gamma_1 & 1
\end{array}
}
Then,  the SU(3) and SU(2) instantons generate the following potential,
\dis{
V= m\Lambda^3_{\rm QCD}\left(\cos(\frac{a}{f_a} +\Gamma_2\frac{a_q}{f_q} )+{\rm h.c.}\right)+f_q^4e^{-2\pi/\alpha_2}\left(\cos(\Gamma_1\frac{a}{f_a}+\frac{a_q}{f_q}) +{\rm h.c.}\right),\label{eq:TwoCos}
}
where $m$ is a typical chiral mass for light quarks, most probably very close to $m_u$. The coefficient in the second term depends on details in the models as shown in \cite{Yosh20,KimSpider09} and we simply denote it by the parameter $f_q^4$.
The mass matrix from Eq. (\ref{eq:TwoCos}) is
\dis{
\begin{pmatrix}
\frac{m\Lqcd^3+\Gamma_1^2 f_q^4 e^{-2\pi/\alpha_2}}{f_a^2},& \frac{\Gamma_1 f_q^4 e^{-2\pi/\alpha_2}+\Gamma_2 m\Lqcd^3  }{f_af_q} \\[0.7em]
 \frac{\Gamma_1 f_q^4 e^{-2\pi/\alpha_2}+\Gamma_2 m\Lqcd^3  }{f_af_q} ,&
 \frac{f_q^4 e^{-2\pi/\alpha_2}+\Gamma_2^2 m\Lqcd^3}{f_q^2} 
\end{pmatrix}. \label{eq:qMassM}
}
The eigenvalues of (\ref{eq:qMassM}) are  
\dis{
m^2_{a,a_q}=&\frac{1}{2 f_a^2 f_q^2} \Biggl(  f_q^4e^{\frac{-2\pi }{\alpha _2}}(f_a^2+f_q^2\Gamma_1^2)+
m\Lqcd ^3 (f_q^2+f_a^2\Gamma_2^2) \\
\\&\mp   \sqrt{-4(\Gamma_1\Gamma_2 -1)^2\,m \Lqcd^3   {f_a^2}{ f_q^6}e^{\frac{-2 \pi }{\alpha _2}} +\left(f_q^4e^{\frac{-2\pi }{\alpha _2}}(f_a^2+\Gamma_1^2f_q^2 ) +m\Lqcd^3(f_q^2+f_a^2\Gamma_2^2)\right)^2 }\,\Biggr).
}
If we expand in terms of $e^{-2\pi /\alpha _2}$ factor,  which is very small, two eigenvalues are
\dis{
m_a^2 &=\frac{m\Lqcd ^3}{ f_a^2 } \Biggl(1+\frac{\Gamma_2^2 f_a^2}{f_q^2}  \Biggr)+O(e^{\frac{-2\pi }{\alpha _2}}  ),\\
m_{a_q}^2&=  (\Gamma_1 \Gamma_2-1)^2f_q^2e^{\frac{-2\pi }{\alpha _2}}   \frac{1}{1+\Gamma_2^2 (f_a^2/f_q^2)} +O(e^{\frac{-4\pi }{\alpha _2}}  ) \, .\label{eq:aMasses}
}
Note that $e^{-2\pi /\alpha _2}$ is almost $0.169\times 10^{-40}$. With $f_q\simeq 10^8 \gev$, we obtain $m_q\simeq 2\times 10^{-13}\gev\approx 0.0002\eV$ and the vacuum energy density $f_q^4 e^{-2\pi /\alpha _2}\approx    (0.64\times 10^{-3}\eV)^4$.

 %%%%%%%%%%%%%%%%%%%%%%%%%%
\subsection{Breaking by potential $V$}\label{subsec:Potential}

Since we try to introduce explicit breaking terms of some U(1) global symmetries, the conditions should come outside those global symmetries that  we considered previously, \USU\,and \USW. Gravitational interactions are known to break any global symmetry, e.g. at order $1/\Mp^n$ as discussed in \cite{Barr92},
\dis{
\frac{\Lambda^{n+4}}{\Mp^n}.\label{eq:GrBreaking}
}
For a given $\Lambda$, the lowest $n$ should be sufficiently large such that a tiny but needed explicit breaking term arises. A possibility for Eq. (\ref{eq:GrBreaking}) to arise for a large $n$ can be a big discrete symmetry. Here we discuss in  the SUSY anti-SU(5) model because it reduces a number of Yukawa couplings of the MSSM and still it is sufficient to present the essence in generating  Eq. (\ref{eq:GrBreaking}) \cite{DKN84,Barr82}. In particular, it is also useful in the discussion of string compactification \cite{LNP954,Kim20ijmpa}. Matter fermions in anti-SU(5) are
\dis{
\ten_g(+1)=\Big(d^c,q,N\Big)_g,~\fiveb_g(-3)=\Big(u^c,\ell \Big)_g,~\one_{g}=(e^+)_g(+5);~g=\{1,2,3\},\label{eq:RepASU5}
}
where quantum number $X=\{1,-3,5\}$ of \anti~is shown in the parantheses, $g$ is the generation index, and the SM doublets are $q$ in $\ten_g(+1)$ and   $\ell$ in $\fiveb_g(-3)$. The Higgs doublets $H_d$ and $H_u$  of the MSSM are included in
\dis{
\five_H(-2)\ni H_d,~\fiveb_H(+2)\ni H_u.
}
We also need the GUT-breaking fields in the anti-SU(5),
\dis{
\Sigma\equiv \ten_H(+1)=\Big(D^c,Q,N'\Big)_H,~\overline{\Sigma}\equiv \tenb_H(-1)=\Big(D,\bar{Q},\bar{N}'\Big)_H\,.\label{eq:Sigmas}
}
We focussed on the anti-SU(5) GUT because it can be obtained at the level-1 in the compactification of the heterotic string \cite{Kim20ijmpa,LNP954}. With the above \anti~GUT representations, the Yukawa coupling for the top quark mass  is allowed at tree level,  $\ten_3\fiveb_3\,\fiveb_H$.  The Yukawa coupling for the down quark mass  is also allowed at tree level,  $\ten_3\ten_3\,\five_H$. However, we must pay attention to the following rare process dimension-5 operators,
\begin{itemize}
\item[(i)] Forbid proton decay \cite{Sakai81,Hall83}: $ \{q_gq_gq_g\ell_g,  d^c_gd^c_gu^c_g N_g\}\in \ten\cdot \ten\cdot \ten\cdot  \fiveb,~  d^c_gu^c_gu^c_g e^+_g\in \ten\cdot \fiveb\cdot  \fiveb\cdot\one$.

\item[(ii)]  Allow  neutrino masses \cite{Weinberg79}: $N' {N}'\ell_g \ell_g H_uH_u\in\ten_H\cdot \ten_H\cdot \fiveb\cdot  \fiveb\cdot \fiveb_H\cdot  \fiveb_H $.

\item[(iii)] Allow a TeV scale $\mu$ term \cite{KimNilles84}: $\mu\,H_uH_d$.
\end{itemize}
 Item (i) has to be sufficiently suppressed but Items (ii) and (iii) must arise. In Table \ref{tab:Anti5}, we considered a few discrete symmetries: matter parity $\Z_2$, R-parity $\Z_{2R}$ and R-parity $\Z_{4R}$. In the Check column Y denotes the satisfied condition and N  denotes the un-satisfied condition. The item (iii) is not satisfied in all  cases with the MSSM fields plus singlets $N$ (in Eq. (\ref{eq:RepASU5})), shown up to the $\fiveb_H$ column,  because the TeV scale of $\mu$ is not a natural one in the MSSM.
%%%%%%%%%
\begin{table}[h!]
\begin{center}
\begin{tabular}{@{}|c| c c ccc|c|c|c|@{}} \hline\\[-1.4em]
 &&& NMSSM& &&\anti &PQ&Check\\
  &$\ten_g$&\quad  $\fiveb_g$ &$\one_g$& $\five_H$&~~$\fiveb_H$   &$\Sigma_{\rm GUT}$ $\overline{\Sigma}_{\rm GUT}$ & $\sigma$ &Yes or No \\[0.2em] \colrule &&&&&&   \\[-1.4em] 
 ~$\Z_2$~  & $+1$  & $+1$ &   $+1$ & $0$& $0$&&  &   (i)\,N,\,(ii)\,Y,\,(iii)\,N    \\[0.3em]
 \\[-1.4em] 
 ~$\Z_{2R}$~  & $+1$  & $+1$ &   $+1$ & $0$& $0$& &  & (i)\,N,\,(ii)\,Y,\,(iii)\,N  \\[0.3em] 
 ~$\Z_{4R}$~  & $+\frac{1}{2}$  & $+\frac{1}{2}$ &   $+\frac{1}{2}$ & $+1$& $+1$  & & & (i)\,Y,\,(ii)\,Y,\,(iii)\,N    \\[0.3em] 
 ~$\Z_{4R}$~  & $+\frac{1}{2}$  & $+\frac{1}{2}$ &   $+\frac{1}{2}$ & $+1$& $+1$& $+4\quad-4$ &$+2$ & (i)\,Y,\,(ii)\,Y,\,(iii)\,Y   \\[0.3em]
 \hline
\end{tabular}
\end{center}
\caption{Discrete symmetries with SUSY anti-SU(5) representations.}\label{tab:Anti5}
\end{table}
%%%%%%%%
%%%%%%%%%
 For  item (iii) to be satisfied, a solution of the gauge hierarchy problem must accompany the discrete symmetry. In this case, we must consider the GUT breaking fields of Eq. (\ref{eq:Sigmas}), where  $\langle\Sigma\rangle= \langle\overline{\Sigma}\rangle=O(10^{17\,}\gev)$ \cite{Ellis89,KimKyae07}. The column PQ $\sigma$ contains the singlet field $\sigma$ whose VEV of O$(10^{10\,}\gev)$ breaks the PQ symmetry to create the invisible axion \cite{KSVZ1}. The  GUT breaking fields allow the following couplings \cite{Ellis89},
 \dis{
 \five_H(-2)\Sigma(1)\Sigma(1)=\epsilon_{abdce}\five^a\ten^{bc}\ten^{de}, ~\fiveb_H(+2)\overline{\Sigma}(-1)\overline{\Sigma}(-1)=\epsilon^{abdce}\five_a\ten_{bc}\ten_{de}.
 }
 So, the colored particle in $\five$ is removed with the anti-colored particle $D^c$ in $\ten$ by $\langle \ten^{45}\rangle $ at the GUT scale. So does  the anti-colored particle in $\fiveb$ with the colored particle $D$ in $\tenb$ by $\langle \tenb_{45}\rangle$. Then, we are left with a pair of Higgs doublets below the scale $\Mg$, and we can consider the following superpotential, generatibng a kind of  $\mu$-term,
\dis{
\ten_{H}^{45}\, \fiveb_{H\,4}\, \fiveb_{H\,5},~\textrm{and }~\tenb^{H}_{45}\, \fiveb^{H\,4}\, \fiveb^{H\,5},
}
which, however, are not present if there is only one pair of the Higgs doublets. If they are present (in case more than one pair of the Higgs doublets are present), then together with the $\five_H\fiveb_H$  term,   the following $\mu$ terms arise
\dis{
W_\mu =\mu_1H_dH_u +\mu_2H_d^1H_d^2+\mu_3H_u^1H_u^2,\label{eq:Bsym}
}
where $\mu_1,\mu_2,$ and $\mu_3$ are  at the GUT scale, leading to the following Higgsino mass matrix,
\dis{~\hskip -0.5cm
&\quad H_d^1\, H_d^2\, H_u^1\, H_u^2\\
\begin{array}{c}
H_d^1\\
H_d^2\\
H_u^1 \\
H_u^2\\
\end{array}
&\begin{pmatrix}
0& \mu_2 &0 &\mu_1\\
\mu_2 &0& \mu_1&0\\
0&\mu_1 &0 &\mu_3 \\
\mu_1&0 &\mu_3&0\\
\end{pmatrix}.
}
With the D-flat condition, leading to $\mu_2=\mu_3$,  four eigenvalues turn out to be
\dis{
\pm (\mu_1 \pm \mu_2).
}
A pair of massless Higgsino arises in case $\mu_2=+\mu_1$ or $-\mu_1$.  Note, however, that $\mu_2=\mu_3=0 $ in Eq. (\ref{eq:Bsym}), as commented above,  due to the bosonic symmetry. At this level, therefore, we cannot derive a pair of massless Higgs doublets. The only way to allow a massless pair is to have $\mu_1=0$, \ie there is no dimension-2 superpotential, as envisioned in \cite{KimNilles84,Munoz93}. This case is shown in the last row of Table \ref{tab:Anti5}, introducing an anti-SU(5) singlet field $\sigma$. Then, a nonrenormalizable term allowed by the discrete symmetry is
\dis{
W\sim \frac{1}{\Mp} \five_H\fiveb_H\sigma\sigma + \five_H(-2)\Sigma(1)\Sigma(1)+\fiveb_H(+2)\overline{\Sigma}(-1)\overline{\Sigma}(-1) 
}
The VEV, $f_a$, of $\sigma$ generates a $\mu=O(\frac{f_a^2}{\Mp})$. If it breaks the PQ symmetry, then it must be at the intermediate scale \cite{Kim84PLB} and the $\mu$ term is at the TeV scale as needed in the MSSM phenomenology \cite{KimNilles84}. This $\sigma$ field may contribute to dark matter in the Universe but not to DE.

 %%%%%%%%%%%%%%%%%%%%%%%%%%
\section{Gravity effects to effective potential}\label{sec:GravCont}

Discrete symmetries are the basis for the appearance of effective global symmetries, which is manifest in the string compactification \cite{Kobayashi07}. 
%%%%%%%%%
\begin{table}[h!]
\begin{center}
\begin{tabular}{@{}|c| c c ccc|c|c|c|@{}} \hline\\[-1.4em]
 &&& NMSSM& &&\anti &PQ&Quintessential\\
  &$\ten_g$&\quad  $\fiveb_g$ &$\one_g$& $\five_H$&~~$\fiveb_H$   &$\Sigma_{\rm GUT},\overline{\Sigma}_{\rm GUT}$ & $\sigma$ &$\sigma_{ q}$ \\[0.2em] \colrule &&&&&&   \\[-1.4em] 
 ~$\Z_{4R}$~  & $+\frac{1}{2}$  & $+\frac{1}{2}$ &   $+\frac{1}{2}$ & $+1$& $+1$  & $+4,\quad+4$ & $r_1=+2$ &    $r_2=+2$  \\[0.3em] 
 ~U(1)$_X$~  & $+1$  & $-3$ &   $+5$ & $-2$& $+2$& $+1,\quad-1$ &$0$ &   $0$ \\[0.3em]
 ~\UPQ~  & $+1$  & $+1$ &   $+1$ & $0$& $0$& $0,\quad~ 0$&  $+1$ &  $\Gamma_2$   \\[0.3em]
 \\[-1.4em] 
 \Uq~  & $+1$  & $+1$ &   $+1$ & $0$& $0$& $0,\quad~ 0$ &  $\Gamma_1$ & $+1$ \\[0.3em] 
 \hline
\end{tabular}
\end{center}
\caption{Quantum numbers of SUSY chiral fields. Heavy quarks in the  KSVZ type axion are not shown.}\label{tab:QNsusy}
\end{table}
%%%%%%%%
Let us illustrate with the $\Z_{4R}$ quantum numbers of the chiral fields in the SUSY anti-SU(5) GUT of Table \ref{tab:Anti5}, shown again in the 2nd row of Table \ref{tab:QNsusy}. Let  the $\Z_{4R}$ be a discrete subgroup of \UoR. The integer quantum numbers $r_1$ and $r_2$ are taken as the \UoR~quantum numbers. Then, an effective superpotential term must carry the \UoR~quantum number $+2$ modulo $4\textrm{ times integer}$.

Consider $\sigma$ first.  For a term $\sigma^n$ to be present, $nr_1=2$ modulo $4\textrm{ times integer}$. For $r_1=1,2,3,4$, the allowed values of $n$ are 2, none, 2, and  none, respectively. So, it is appropriate to choose $r_1=2$ or $ 4\textrm{ (times integer)}$ so that the PQ symmetry breaking is soley through the \UPQ-SU(3)$_{\rm color}^2$ anomaly.
 But, if we consider $\five_H\fiveb_H$ also, we must choose $r_1=2$ such that $\five_H\fiveb_H\sigma$ is forbidden.
 
Next, consider $\sigma_q$.   Again, to allow a term $\sigma_q^n$, one can choose   $nr_2=2$ or $ 4\textrm{ (times integer)}$ so that the weak-SU(2) symmetry breaking is soley through the \Uq-SU(2)$_W^2$ anomaly.
With a minimal K$\ddot{\rm a}$hler potential term, a superpotential term $W\sim \sigma_q^{n+3}/\Mp^n$ would allow  the vacuum energy of order $V\sim (n+3)|\langle\sigma_q\rangle|^{n+3}/\Mp^{n-1}$.
If we take $\langle\sigma_q\rangle=1.48\times 10^8\gev$ as implied by Eq. (\ref{eq:SU2scale}), we must forbid the terms in the superpotential up to the order
\dis{
\frac{\sigma_q^{11.82}}{\Mp^{8.82}},
}
which needs the superpotential terms for $n\ge 9$. But,
with the aforementioned value for $n=2$ or 4, we cannot forbid terms with $n\le 8$.  $\sigma_q$ alone cannot have a solution.

Therefore, let us consider the other fields of Table \ref{tab:QNsusy} also, especially $\five_H, \fiveb_H, \Sigma_{\rm GUT}$, and $ \overline{\Sigma}_{\rm GUT}$. If we choose $n=4$ (or 2) for $\sigma_q$, the superpotential $\five_H\fiveb_H\sigma_q$ is (or not) allowed. So, let us choose $r_2=2$ for $\sigma_q$. In addition, inclusion of $\Sigma$ and $\overline{\Sigma}$ is required not to allow the above result, and hence $\Z_{4R}$ quantum numbers of  $\Sigma$ and $\overline{\Sigma}$ are $\pm 4$. Then, the GUT scale VEVs of these fields do not break the  $\Z_{4R}$ symmetry, as required. These $\Z_{4R}$ quantum numbers are given in Table \ref{tab:QNsusy}.

For $\Gamma_1\Gamma_2=1$, viz. Eq. (\ref{eq:aMasses}), the QA mass is only through the terms from the superpotential.  The  $\Z_{4R}$ allowed terms are

\dis{
\sigma^2\sigma_q,~\sigma\sigma_q^2,~\Sigma\overline{\Sigma}\sigma ,~\Sigma\overline{\Sigma}\sigma_q ,~\five_H\fiveb_H\sigma^2, \five_H\fiveb_H\sigma_q^2, ~\rm etc.
}
In particular,  we need the following for the electroweak scale $\mu$-term \cite{KimNilles84},
\dis{
\frac{1}{\Mp}\five_H \fiveb_H\sigma^2.
}
But the terms $\sigma^2\sigma_q,~\sigma\sigma_q^2,~\Sigma\overline{\Sigma}\sigma ,$ and $\Sigma\overline{\Sigma}\sigma_q $                                      
should be forbidden. It can be achieved by considering $\Z_{nR}\,(n\ge 9)$ or $\Z_{4R}\times \Z_{\rm matter}.$       For  $\Z_{4R}\times \Z_{2}$, the quantum numbers are shown in Table \ref{tab:QNtwoZs}.

 %%%%%%%%%
\begin{table}[h!]
\begin{center}
\begin{tabular}{@{}|c| c c ccc|c|c|c|@{}} \hline\\[-1.4em]
 &&& NMSSM& &&\anti &PQ&Quintessential\\
  &$\ten_g$&\quad  $\fiveb_g$ &$\one_g$& $\five_H$&~~$\fiveb_H$   &$\Sigma_{\rm GUT},\overline{\Sigma}_{\rm GUT}$ & $\sigma$ &$\sigma_{ q}$ \\[0.2em] \colrule &&&&&&   \\[-1.4em] 
 ~$\Z_{4R}$~  & $+\frac{1}{2}$  & $+\frac{1}{2}$ &   $+\frac{1}{2}$ & $+1$& $+1$  & $+4,\quad+4$ & $r_1=+2$ &    $r_2=+2$  \\[0.3em] 
 ~$\Z_{2}$~  & $0$  & $0$ &   $0$ & $+1$& $+1$& $+1,\quad+1$ &$+1$ &   $+1$ \\[0.3em]
 \hline
\end{tabular}
\end{center}
\caption{Working quantum numbers of SUSY chiral fields. Not only one but a few heavy quarks (because the SM quarks carry PQ charges) are assumed to be added to make the domain wall number one.}\label{tab:QNtwoZs}
\end{table}
%%%%%%%%
  
%%%%%%%%%%%%%%%%
\section{Conclusion}\label{sec:Conclusion}
We discussed the possibility that the weak SU(2) gauge symmetry works for the source of explicit breaking terms of a global symmetries suggested for a quintessential axion. We find that it is possible in supersymmetric standard models if the vacuum expectaion value of the quintessential axion field is  about $1.48\times 10^{8\,}\gev$.
 
 %%%%%%%%%%%%%%%%%%%%%%%%%%%%%%%%%%%%%%%%%%%%%%%%%%%%
\acknowledgments{I have indebted from Junu Jeong and Myungbo Shim for checking the needed numbers. This work is supported in part by the National Research Foundation (NRF) grant  NRF-2018R1A2A3074631.} 

  %%%%%%%%%%%%%%%%%%  

\end{document}